\documentclass{article}

\usepackage{arxiv}

\usepackage[utf8]{inputenc} 
\usepackage[T1]{fontenc}    
\usepackage{hyperref}       
\usepackage{url}            
\usepackage{booktabs}       
\usepackage{amsfonts}       
\usepackage{nicefrac}       
\usepackage{microtype}      
\usepackage{lipsum}		
\usepackage{graphicx}
\usepackage{doi}

\title{Exploring Low Cost Non-Contact Detection of Biosignals for HCI}

\author{ \href{https://orcid.org/0000-0000-0000-0000}{\includegraphics[scale=0.06]{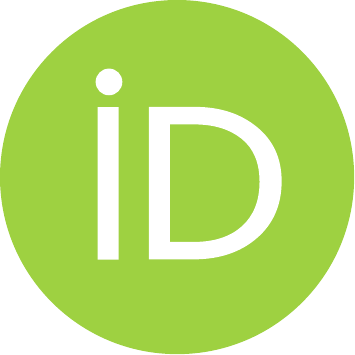}
\hspace{1mm}Christoph Tremmel} \\
	School of Electronics and Computer Science\\
	University of Southampton\\
	Southampton, UK \\
	\texttt{christoph.tremmel@southampton.ac.uk} \\
	\And
	\href{https://orcid.org/0000-0000-0000-0000}{\includegraphics[scale=0.06]{orcid.pdf}
 \hspace{1mm}Indu Bodala} \\
	School of Electronics and Computer Science\\
	University of Southampton\\
	Southampton, UK \\
	\texttt{I.P.Bodala@soton.ac.uk} \\
	\And
	\href{https://orcid.org/0000-0000-0000-0000}{\includegraphics[scale=0.06]{orcid.pdf}
 \hspace{1mm}m.c. schraefel} \\
	School of Electronics and Computer Science\\
	University of Southampton\\
	Southampton, UK \\
	\texttt{mc@ecs.soton.ac.uk} \\
}

\renewcommand{\shorttitle}{\textit{arXiv} Template}

\hypersetup{
pdftitle={Exploring Low Cost Non-Contact Detection of Biosignals for HCI},
pdfsubject={q-bio.NC, q-bio.QM},
pdfauthor={Christoph Tremmel, Indu Bodala, m.c. schraefel},
pdfkeywords={remote vital sign detection, human computer interaction, remote photoplethysmography, heart rate, respiration rate},
}

\begin{document}
\maketitle

\begin{abstract}
In an effort to make biosignal integration more accessible to explore for more HCI researchers, this paper presents our investigation of how well a standard, near ubiquitous webcam can support remote sensing of heart rate and respiration rate across skin tone ranges. The work contributes: how the webcam can be used for this purpose, its limitations, and how to mitigate these limitations affordably, including how the skin tone range affect the estimation results. 
\end{abstract}


\section{Introduction}
This paper presents our exploration into the feasibility of using standard web cameras to support contactless biosignal detection as part of potentially richer virtual interactions. HCI research has a long history of interest and use of biosignal detection, particularly heart rate (HR) \cite{ dey2017effects, slovak2012understanding}, heart rate variablity (HRV) \cite{rowe1998heart, mcduff2016cogcam} and respiration rate (RR) \cite{moraveji2012breathtray}.
Uses for these signals in HCI include exertion gaming, such as sharing HRs to run together apart \cite{mueller2010jogging}.
Biofeedback is another application. Hassib and colleagues \cite{hassib2017heartchat} presented an HR-augmented mobile chat app that used HR as a proxy of empathy between users, as well as to promote engagement and play in chat activity.
Respiration rate is often used to slow breathing for stress reduction \cite{magnon2021benefits}. This was demonstrated by Harris and colleagues\cite{harris2014sonic} when they used participants' RR to adjust the quality of music. Moraveji et al. \cite{moraveji2011peripheral} showed they could reduce the breathing rate of participants with peripheral stimuli such as a moving or dimming desktop computer menu. In sum, these signals - pulse and respiration  - are well known to be associated with both affective \cite{riener2009heart,magagnin2010heart}, attentive \cite{porges1969respiratory, ramirez2015anxiety} and cognitive \cite{forte2019heart, varga2017rhythms} states. In HCI, as per each of the above examples, exploration of these signals for interaction has been constrained by the requirements  to use purpose-built devices for signal detection. That constraint limits the opportunities of designers and researchers to explore how biosignals may be used more diversly and effectively in interactions.

Related work shows that at least HR and RR can be detected remotely, mainly by a camera looking at one's face \cite{mcduff2021camera, hwang2021non}. Given the degree to which both work and social communication are carried out via webcam-based virtual meetings, this seems a rich space to explore biosignal integration. For example, in our lab, we have been looking at how richer co-awareness of personal state informed by biosignals during virtual meetings may contribute to helping such meetings feel more like in-person encounters. 

Since webcams are now near ubiquitous, it seems a clear win to be able to make use of this camera-as-sensor to capture these biosignals in these contexts. The main question is then: are these devices up to the task? In this paper, we present our work to interrogate this question. In the following sections, we overview the related work in contactless remote biosignal detection. We overview the main video techniques that can be used for remote HR and RR, followed by presenting how tested these techniques. We then discuss our results in terms of usefulness for HCI.

There are two main contributions of this study:
1. \textit{Webcams for Remote Biosignal Detection}. Our main goal has been to open up access to and use of biosignals for HCI designers and researchers to explore in as accessible/affordable a way as possible. Inspired by Buxton's taxonomy of input devices \cite{buxton1983lexical}, our results provide an overview of the affordances and constraints of lightweight webcam use for biosignal detection, so that the HCI community can make decisions about deployment by understanding the constraints and affordances of these devices. 
2. \textit{Addressing a key inclusivity limitation}. In the techniques described below, skin tone is an established factor in detection using the main established methods. As part of our investigation, we have paid particular attention to this feature, and have tested how inclusive remote heart rate estimation really is by exposing it to variety of skin tones.

\section{Related Work}
In the introduction we identified that the typical HCI approach to  capture biosignals requires first, specialised devices and second, require contact with the participant. In this section we characterise related work in contactless biosignal detection. 
Publications involving remote detection of biosignals like HR, RR, pulse oxymetry and gaze detection have been growing nearly exponentially in the last 10 years \cite{mcduff2021camera} particularly in the medical field to be able to support remote, contactless monitoring of vital signals, for example, during remote consultations \cite{annis2020rapid} or neonatal monitoring \cite{aarts2013non}.
We suggest there is a significant opportunity for HCI to leverage this work towards less clinical applications. As context for our work, in the following, we review related work in HCI and then overview the main techniques for remote detection of biosignals.

\subsection{Biosignal Detection in HCI }

 From our exploration of the related work in HCI specifically, there seem to be only three ways  that remote/contactless biosignal detection has been used, and and only to use one biosignal, HR.
1. \textit{Task Switching}  McDuff et al \cite{mcduff2016cogcam} in their paper
 \textit{Features from HR for Cognitive Task Detection} use remote photoplethysmography (rPPG), described below, to differentiate between multiple tasks based on multiple cognitive stress features extracted from these readings. Although it has to be noted that there was no report on the accuracy of the HR estimates. Likewise the camera used was more sophisticated than a web cam.
2. Biological signals displayed as \textit{Biofeedback} for participants to guide responses to change in biosignals. This biofeedback approach in  has been applied by Frey \cite{frey2016remote} in a board game-like settings and and the reactions of the players to the feedback has been documented. 
3. \textit{Passive BioData Capture} Poh et al.\cite{poh2011medical} proposes a "medical mirror" that can acquire HR readings remotely that, they suggest, could easily be integrated into the everyday life, but based on our searches, this has not yet been implemented.

We could not find any publications in the HCI literature with remote respiratory detection. Examples outside HCI are described in the section below, in our overview of detection mechanisms.
\subsection{Detecting HR and RR remotely}
In this section, we overview contactless HR and RR detection as it informs our focus on remote detection via webcams. 

\textbf{Heart Rate (HR)} is the frequency of heart contractions usually measured in beats per minute while HRV describes the variation of the interval between contractions. Both are  significant measures of human state and are well associated with a multitude of health symptoms and body functions. HR for example increases as a response to stress, physical workout, drug use etc. An increased resting HR also correlates with mortality \cite{palatini2007heart}. Meanwhile a decreased HRV is, for example, associated with increased work stress, physical fatigue, increased morbidity and mortality \cite{thayer2010relationship}. Both measures are usually acquired using electrocardiogram (ECG), sphygmomanometry or photoplethysmography (PPG), all of which require technical equipment, contact to the human skin, and in a clinical setting, often trained personnel while exposing patients to uncomfortable measuring sessions. Recently, consumer grade equipment like smart watches and pulse oxymeters have gained popularity since they can acquire HR easier and without external assistance, however these devices still require close skin contact.

\textbf{Respiratory rate (RR)} is the frequency in which breathing occurs and is usually measured in breaths per minute. While it is a vital sign that is the least often recorded it can still hold strong value for certain measurements, as, for example, a predictor of cardiac arrest \cite{cretikos2008respiratory}. This measure can be acquired manually by counting each individual breath, spirometry, capnometry or PPG among others.

\subsubsection{Remote Heart Rate Detection}
\label{remoteHR}
Plainly, moving away from direct contact with the skin introduces a number of challenges to achieve similar signal quality to regular contact-based PPG. There are several techniques for remote tracking of HR such as  sound \cite{zhang2010sound}, near-infrared \cite{vogels2018fully}, far-infrared/thermal\cite{gault2013fully}, ultrasound \cite{hamelmann2019doppler}, radar \cite{li2013review}, ballistocardiography (BCG) \cite{starr1939studies} or remote photoplethysmography (rPPG) \cite{gudi2020real}. Among these techniques rPPG seems to be the most promising as it can be utilized using everyday cameras such as digital photography cameras, webcams or smartphone cameras. It also does not require to be completely motionless as it is necessary with camera based BCG, which measures HR through the motion of the human body caused by the heart beat. RPPG on the other side measures light that has been reflected from the skin and captures the changes in blood flow in the micro-vascular tissue bed underneath the skin caused by the heart beat \cite{allen2007photoplethysmography}. Challenges to overcome in devices that are not purposely built for rPPG like standard RGP cameras include related factors from illumination to lens size. Perhaps most critically, rPPG requires suitable illumination, a close distance between camera and user, as well as a limited amount of movement \cite{shirbani2020effect, hassan2020towards}. However, motion artifact reduction has also improved significantly \cite{cennini2010heart,feng2014motion}. It is also important that the algorithm always selects the same region of interest (ROI). Most of the algorithms use the face or parts of the face as ROI and therefore employ face tracking algorithms like the Viola-Jones algorithm. Failure to track the ROI because of movement or concealed users (by hats or other accessories) will result in incorrect signals \cite{hassan2020towards}. Additionally, the ROI has to be selected carefully so that hair, beards or headgear do not partially cover the area \cite{stricker2014non, kwon2015roi} and users have to be instructed to not wear make-up or any other product that might interfere with the camera \cite{wang2020impact}. Finally, since rPPG is using colour changes to detect the HR, it is affected by the skin tone of the user. A darker skin colour absorbs more and reflects less light and is therefore associated with poorer detection \cite{hassan2020towards}. From the hardware side it is important to note, that webcams themselves are usually implemented to automate exposure, focus and image compression. These automatic features negatively affect the estimation, though, in some cases can be switched off. Also, the cameras' physical lens size and frequent limitations to just RGB are also challenges, though not insurmountable, for successful reading. \cite{poh2010advancements, gudi2020real}. From the algorithm side there are two main approaches: signal processing methods and supervised methods. Former are based on traditional signal processing methods and are easier to implement and understand, and require no training data. Supervised methods are most of the time based on convolutional neural networks or transformers. While they require a lot of training data and are usually more computational they have been shown to separate noise from the desired HR more effectively. Ultimately it can be said that, in good conditions like little movement, close distance, good lighting and a camera that doesn't alter the video too much, HR estimates with very low error are possible.

\subsubsection{Remote Respiration Rate Detection}
Similar to remote HR tracking, there are many different approaches to remotely detect the respiratory rate such as techniques based on radar \cite{xiao2006frequency, choi2009remote}, interferometry \cite{mikhelson2011remote}, temperature \cite{basra2017temperature}, humidity \cite{xiao2019fast}, ultrasound \cite{min2007study}, rPPG \cite{van2016robust} or camera \cite{zhao2013remote}.
While lidar and related approaches are finding their way into higher end mobile devices, mainly to support AR \cite{gil2013use}, for personal use involving only everyday equipment only rPPG and camera based techniques are currently viable as ubiquitous detection technology. As already established rPPG is a remote technique to measure HR and breathing modulates the acquired rPPG waveform in three ways: intensity, frequency and amplitude. Six respiratory-induced variations such as period of pulse have been identified to correlate with the respiratory rate \cite{li2010comparison}. This method is quite sensitive as it requires a robust rPPG signal to work which requires near perfect conditions (good lighting, no movement, no automated camera function etc. as described in \ref{remoteHR}) to produce good results which can be quite challenging to apply in real-life situations. The camera approach uses the movement of the chest and shoulders during respiration to extract the respiratory rate \cite{zhao2013remote, hwang2021non}. Since these movements are quite large compared to BCG it easier to detect but still very susceptible to any artifacts generated by body motions. Similarly to rPPG, the two main approaches for algorithms are divided into signal processing and supervised learning. Since colour detection is not necessary for this method it can also be achieved during low light conditions particularly when combined with near-infrared light sensors \cite{zhao2013remote}. Ideally, this approach should be used in static activities such as working at computer (including our virtual meeting scenarios), watching TV or sleeping. 

\section{Assessment of Webcam for HR/RR Detection }

In order to evaluate how well a standard webcam can acquire HR and RR we developed an experimental setup that approximates typical activities that are performed in front of a computer. For this we split our experiment into two parts where in the first half we developed guided eye tasks based on standardized eye movements, including saccades \cite{purves2019neurosciences} and smooth pursuits \cite{purves2019neurosciences}. The second half introduced different activities related to breathing and HR such as holding the breath or increased RR/HR due to physical activity. Combined these tasks offered us a repeatable way to simulate physical aspects of work at a computer likely to influence HR and RR - letting us assess a dynamic range in signal capture. In the following section we discuss our study protocol.

\subsection{Experimental Setting and Signal Acquisition}
The Participants for this study were recruited among students and employees of the University of Southampton. Ethical approval was received on the 12th of May 2022 (Application ID XX). We ensured that participants with a  wide variety of skin tones and racial backgrounds were included.
Each experimental session lasted roughly one hour including preparation time. In total 18 participants participated in this study. Two participants had to be excluded from the results as one subject was wearing make-up making the HR estimation impossible and another had open hair which confused the facial detection. That leaves 16 in total (10 males, mean age 30.75, SD 5.95).
Participants sat comfortably during the whole study in front of an LCD screen. An Aputure AL-H198C LED (11x18 grid pattern) was positioned behind the monitor set at 3430 Lux, 3200 Kelvin to ensure the same lighting conditions and reduced influence of ambient light for each participant. For this experiment electrocardiogram (ECG), electrooculogram (EOG) and respiratory rate were being acquired at 128 Hz by using a Mindmedia NeXus-10 MKII biomedical amplifier to serve as ground truth signals. One-time use sticky electrodes were placed at the inside of each arm for ECG, right/left of both eye for horizontal EOG and above/below the left eye for vertical EOG. For RR a respiration belt was strapped on the participant. A Logitech Brio camera was used to record the subjects at 30 Hz with full HD resolution (1920x1080). In order to ensure synchrony between the acquired signals trigger signals were sent from the PC that displayed the experiment tasks to the amplifier via a parallel port. The trigger to camera started the video recording for a fix number of frames that were saved to the PC's RAM during the task and saved to disk in the wait time between the trials. This procedure ensured a stable frame rate that is essential to compare the video to the physiological signals. For the following analysis, only ECG, respiratory data and video recordings were included.

\subsection{Experimental Task}
The experiment consisted of two parts: one part focused mostly on respiratory exercises while the second focused on visual object tracking. The order of both parts was alternated between each participant. The respiratory part \emph{which was conducted without any EOG electrodes} was split again into two parts where the first part was initiated at sitting/resting heart rate, and the second part was after a set of exercises to elevate heart rate. The workout consisted of walking up five floors (~3 mins activity) to increase the subjects' HR and RR. The experiment had a total of seven tasks:
\begin{itemize}
 \item[]1) Breath normally
 \item[]2) Hold Breath: Participants had to hold their breath for the time, if for any reason they were out of air they were encouraged to resume breathing
 \item[]3) Horizontal saccadic eye movements: Participants had to follow a saccadic flashing object that sped up over the course of the trial from 2/3 to 4 Hz.
 \item[]4) Vertical saccadic eye movements
 \item[]5) Horizontal smooth eye movements: A grid of grey 8x6 crosses would be displayed and participant had to follow the red-highlighted cross with their eyes. The speed of the highlighted cross was 4 Hz.
 \item[]6) Vertical smooth eye movements
 \item[]7) Diagonal smooth eye movements
\end{itemize}

Each block starts with one second of preparation where the display shows the next task and the word "ready", then the task is presented. After the task, while the recorded video is saved to disk, the display shows the next task and the word "Wait".
The respiratory part included 5 blocks for each condition (normal/workout) where each block consisted of a randomly permuted order of tasks 1, 2 and 7 with a trial length of 20 seconds. The total recording time for this part was therefore 3 (tasks) * 20 (trial length) * 5 (blocks) * 2 (conditions) = 600 seconds.
The eye movement part included 10 blocks where each block consisted of a randomly permuted order of task 3-7 with a trial length of 10 seconds. The total recording time for this part was 5 (tasks) * 10 (trial length) * 10 (blocks) = 500 seconds. 

\section{Methods}
For HR and RR we calculated the estimate for one trial and reported the root mean square error between estimate and real signal. As a first step, in order to reduce the amount of saved data, each frame was cropped. 300 pixels from each side and 200 pixels from the top were removed which resulted in a resolution of 1320 x 880. Afterwards, we applied the classical Viola Jones face tracking \cite{viola2001rapid} on each frame. Then the processing pipeline differs for each estimate.

For HR we selected the lower half of the face as ROI. We deliberately did not select the forehead as hair can easily produce artefacts. After feature extraction using the spherical mean \cite{pilz2019vector} method, the signal was filtered between 0.7 and 2.5 Hz. Then, the frequency with the greatest power was calculated by applying a short-time Fourier transform on a moving window for a trial. Finally, the median frequency over all windows was reported as estimate for each trial. The HR ground truth signal was calculated by first applying a peak detection on the first derivation of the baseline corrected ECG signal and then using the peaks with a spline interpolation to create a PPG-like signal. The frequency was then calculated exactly the same as for the estimate. 

For the RR estimate we selected the area below the face until the bottom of the camera as ROI similar to \cite{zhao2013remote}. Participants were positioned so that shoulders and chest were still in the picture. We then converted the ROI to grayscale and used the average pixel as feature. The signal was then filtered between 0.2 and 0.5 Hz. We calculated the respiratory rate estimate and ground truth in a similar manner as the HR estimate by applying a short-time Fourier transform on a moving window to get the greatest power frequency and used the median frequency as the result for each trial. We did not use any blind-source separation techniques like independent component analysis as we were instructing participants to keep movement to a minimum. 

For supplementary information on the ROI and how typical signals looked like we refer to the appendix figure \ref{fig:ROI} and \ref{fig:estimate}. 

\section{Results}
\begin{figure}
    \centering
    \begin{minipage}{0.45\textwidth}
        \centering
        \includegraphics[width=1\textwidth]{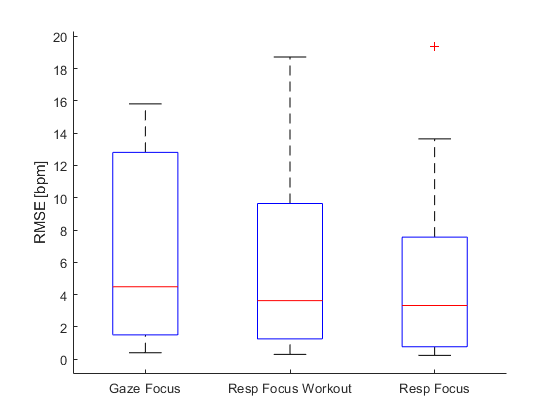} 
        \caption{Boxplots for the HR prediction RMSE averaged over all participants for the three conditions. The respiration conditions shows the lowest error, then the workout condition and finally the gaze condition.}
        \label{fig:Hearterror}
    \end{minipage}\hfill
    \begin{minipage}{0.45\textwidth}
        \centering
        \includegraphics[width=1\textwidth]{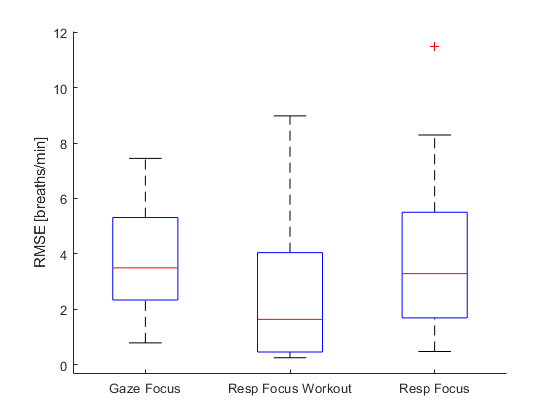} 
        \caption{Boxplots for the RR prediction RMSE averaged over all participants for the three conditions. The workout condition shows the lowest error, the other two conditions perform similar.}
        \label{fig:resperror}
    \end{minipage}
\end{figure}

Figure \ref{fig:Hearterror} shows the boxplots for heat rate prediction RMSE averaged over all participants. It can be seen that median error is between 3-5 bpm depending on the condition. The gaze focus condition has the highest median error with 4.48, the workout condition comes in the middle with 3.62 and then respiratory focus without work has the lowest error with 3.32. It should also be noted that the 75th percentile measure for the gaze focus condition is a lot higher than for the other two conditions indicating a higher variance of estimation quality.

\begin{figure}[h]
\centering
\includegraphics[scale=0.7]{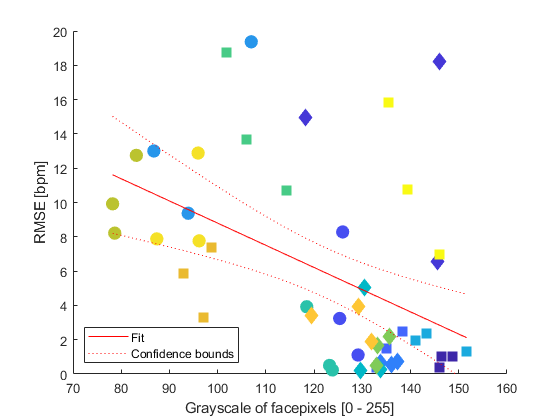}
\caption{Each marker shape and colour combination represents one participant for each of the three conditions and shows the relationship between the skin colour in grayscale and the RMSE between prediction on ground truth. The red line shows a linear fit with confidence bounds.}
\label{fig:skinclourerror}
\end{figure}

In figure \ref{fig:skinclourerror} the relationship between the estimation error and the skin colour is shown. Each participant is portrayed by a unique colour/shape combination and has three data points representing the three experiment conditions. The skin colour is hereby displayed as the to-grayscale converted face pixels of each participant. We see clearly that the higher the grayscale, i.e. the lighter the face, the lower the error. This correlation is also shown by the a linear fitting line. While this result could also be affected by ambient light, in this assessment,  the light conditions for each experiment were the same. It can be seen that, but for one outlier, the grayscale values across participants are clustered closely to each other. This proximity indicates only little colour difference such as based on ambient light during the recording sessions.


Plot \ref{fig:resperror} displays the boxplots for respiratory rate prediction RMSE averaged over all participants. Task 2 "Hold Breath" was excluded from this result. It can be seen that the median error is between 1-4 breaths per minute. The lowest error was achieved in the workout condition with 1.64, while the other two conditions perform equally well with 3.29 for the respiration focus and 3.5 for the eye gaze focus.

\section{Discussion}
From the above data we can see the following. For HR the median RMSE between prediction and groundtruth was 3-5 bpm depending on the setting. The shorter time windows (10 sec vs 20 sec) and the EOG electrodes in participants' faces during the gaze focus experiment made the prediction harder and therefore come with the worst results. Then come the results for the workout condition. Here it can be seen that the median result is nearly identical to the non-workout condition but there are more participants with a higher error. This makes sense as some participants tended to breath heavier and therefore move more which introduced more artifacts. The best condition was the non-workout condition where participants nearly sat motionless especially during the 'hold you breath' trials. Figure \ref{fig:skinclourerror} shows that there is a clear influence of darker skin colour on the estimation error. The grayscale of skin pixels was used as an approximation of skin colour since identical lighting conditions for participants were used. The plot shows that estimation of light skin colour can nearly be done perfectly, while darker skin colours are off by up to 10 bpm. These results are in a similar range of what other publications report especially when it is not clear what the skin colour of participants were \cite{gudi2020real, wang2018comparative, hassan2020towards, dasari2021evaluation}. We have noticed that the Logitec Brio is not the ideal camera for these types of methods as there seems to be an internal compression that reduces the image quality. Nevertheless, our work shows is that even with compression - which most webcams will have as default - the results are still usable because of the low estimation error. A camera without compression could lead to even better results and manage to compensate for skin colour.

The results for the respiratory rate detection were between 1-4 breaths per minute depending on the condition. Here the workout condition clearly performed best because participants were taking deeper breaths after physical effort, thus making larger movement that could be detected more easily. The estimation error for the other conditions were nearly identical. Altogether, these results are only slightly worse than in \cite{zhao2013remote} but critically, we note that in that study, a night vision camera was used which was superior (bigger lens size, lossless recording) to our, and indeed most, standard webcams.

\subsection{Take Aways}
Our results show that with a basic webcam and appropriate conditions (good lighting, little movement, close distance to camera) it is possible to estimate HR and RR with decent accuracy using relatively short windows compared to other studies. This also indicates that real time acquisition of these signals is well possible.
To further improve these results especially in respect to people with darker skin colour, we recommend a webcam that does not employ autoexposure/focus functions and saves the video stream without or with only very little compression (such as the Logitech C925e). 

\section{Conclusion}
In this paper we present preliminary results for remote HR and RR detection using a common webcam, assessed via different experimental tasks. Our goal has been to evaluate the effectiveness of this near-ubiquitous input device, first, to enable more HCI researchers to explore integrating remote biosignal detection into their work, and second, to ensure better inclusivity of different skin tones. 
The paper offers several contributions. First it shows that current webcams \textit{can} be used for HR/RR detection, and provides a set of limitations for consideration in an application. Second it identifies the factors that mitigate these limitations. Third, it shows that, for relatively low cost, researchers can get webcams that offer these limiting parameters as controllable, out-of-the-box, enabling far better results in detection of HR and RR, in terms of precision, and especially across skin colouring. 
With these results, HCI researchers can see they can now add remote biosignal detection in their design explorations, determine if they require devices with greater fidelity than what they have on their desktop already, and if so, know which parameters need to be adjustable.


\bibliographystyle{abbrv}
\bibliography{references}

\begin{thebibliography}{10}

\bibitem{aarts2013non}
L.~A. Aarts, V.~Jeanne, J.~P. Cleary, C.~Lieber, J.~S. Nelson, S.~B. Oetomo,
  and W.~Verkruysse.
\newblock Non-contact heart rate monitoring utilizing camera
  photoplethysmography in the neonatal intensive care unit—a pilot study.
\newblock {\em Early human development}, 89(12):943--948, 2013.

\bibitem{allen2007photoplethysmography}
J.~Allen.
\newblock Photoplethysmography and its application in clinical physiological
  measurement.
\newblock {\em Physiological measurement}, 28(3):R1, 2007.

\bibitem{annis2020rapid}
T.~Annis, S.~Pleasants, G.~Hultman, E.~Lindemann, J.~A. Thompson, S.~Billecke,
  S.~Badlani, and G.~B. Melton.
\newblock Rapid implementation of a covid-19 remote patient monitoring program.
\newblock {\em Journal of the American Medical Informatics Association},
  27(8):1326--1330, 2020.

\bibitem{basra2017temperature}
A.~Basra, B.~Mukhopadhayay, and S.~Kar.
\newblock Temperature sensor based ultra low cost respiration monitoring
  system.
\newblock In {\em 2017 9th International Conference on Communication Systems
  and Networks (COMSNETS)}, pages 530--535. IEEE, 2017.

\bibitem{buxton1983lexical}
W.~Buxton.
\newblock Lexical and pragmatic considerations of input structures.
\newblock {\em ACM SIGGRAPH Computer Graphics}, 17(1):31--37, 1983.

\bibitem{cennini2010heart}
G.~Cennini, J.~Arguel, K.~Ak{\c{s}}it, and A.~van Leest.
\newblock Heart rate monitoring via remote photoplethysmography with motion
  artifacts reduction.
\newblock {\em Optics express}, 18(5):4867--4875, 2010.

\bibitem{choi2009remote}
J.~H. Choi and D.~K. Kim.
\newblock A remote compact sensor for the real-time monitoring of human
  heartbeat and respiration rate.
\newblock {\em IEEE transactions on biomedical circuits and systems},
  3(3):181--188, 2009.

\bibitem{cretikos2008respiratory}
M.~A. Cretikos, R.~Bellomo, K.~Hillman, J.~Chen, S.~Finfer, and A.~Flabouris.
\newblock Respiratory rate: the neglected vital sign.
\newblock {\em Medical Journal of Australia}, 188(11):657--659, 2008.

\bibitem{dasari2021evaluation}
A.~Dasari, S.~K.~A. Prakash, L.~A. Jeni, and C.~S. Tucker.
\newblock Evaluation of biases in remote photoplethysmography methods.
\newblock {\em NPJ digital medicine}, 4(1):1--13, 2021.

\bibitem{dey2017effects}
A.~Dey, T.~Piumsomboon, Y.~Lee, and M.~Billinghurst.
\newblock Effects of sharing physiological states of players in a collaborative
  virtual reality gameplay.
\newblock In {\em Proceedings of the 2017 CHI conference on human factors in
  computing systems}, pages 4045--4056, 2017.

\bibitem{feng2014motion}
L.~Feng, L.-M. Po, X.~Xu, Y.~Li, and R.~Ma.
\newblock Motion-resistant remote imaging photoplethysmography based on the
  optical properties of skin.
\newblock {\em IEEE Transactions on Circuits and Systems for Video Technology},
  25(5):879--891, 2014.

\bibitem{forte2019heart}
G.~Forte, F.~Favieri, and M.~Casagrande.
\newblock Heart rate variability and cognitive function: A systematic review.
\newblock {\em Frontiers in neuroscience}, 13:710, 2019.

\bibitem{frey2016remote}
J.~Frey.
\newblock Remote heart rate sensing and projection to renew traditional board
  games and foster social interactions.
\newblock In {\em Proceedings of the 2016 CHI Conference Extended Abstracts on
  Human Factors in Computing Systems}, pages 1865--1871, 2016.

\bibitem{gault2013fully}
T.~Gault and A.~Farag.
\newblock A fully automatic method to extract the heart rate from thermal
  video.
\newblock In {\em Proceedings of the IEEE conference on computer vision and
  pattern recognition workshops}, pages 336--341, 2013.

\bibitem{gil2013use}
E.~Gil, J.~Llorens, J.~Llop, X.~F{\`a}bregas, and M.~Gallart.
\newblock Use of a terrestrial lidar sensor for drift detection in vineyard
  spraying.
\newblock {\em Sensors}, 13(1):516--534, 2013.

\bibitem{gudi2020real}
A.~Gudi, M.~Bittner, and J.~van Gemert.
\newblock Real-time webcam heart-rate and variability estimation with clean
  ground truth for evaluation.
\newblock {\em Applied Sciences}, 10(23):8630, 2020.

\bibitem{hamelmann2019doppler}
P.~Hamelmann, R.~Vullings, A.~F. Kolen, J.~W. Bergmans, J.~O. van Laar,
  P.~Tortoli, and M.~Mischi.
\newblock Doppler ultrasound technology for fetal heart rate monitoring: a
  review.
\newblock {\em IEEE transactions on ultrasonics, ferroelectrics, and frequency
  control}, 67(2):226--238, 2019.

\bibitem{harris2014sonic}
J.~Harris, S.~Vance, O.~Fernandes, A.~Parnandi, and R.~Gutierrez-Osuna.
\newblock Sonic respiration: controlling respiration rate through auditory
  biofeedback.
\newblock In {\em CHI'14 Extended Abstracts on Human Factors in Computing
  Systems}, pages 2383--2388. 2014.

\bibitem{hassan2020towards}
M.~Hassan, A.~Malik, D.~Fofi, B.~Karasfi, and F.~Meriaudeau.
\newblock Towards health monitoring using remote heart rate measurement using
  digital camera: A feasibility study.
\newblock {\em Measurement}, 149:106804, 2020.

\bibitem{hassib2017heartchat}
M.~Hassib, D.~Buschek, P.~W. Wozniak, and F.~Alt.
\newblock Heartchat: Heart rate augmented mobile chat to support empathy and
  awareness.
\newblock In {\em Proceedings of the 2017 CHI Conference on Human Factors in
  Computing Systems}, pages 2239--2251, 2017.

\bibitem{hwang2021non}
H.-S. Hwang and E.-C. Lee.
\newblock Non-contact respiration measurement method based on rgb camera using
  1d convolutional neural networks.
\newblock {\em Sensors}, 21(10):3456, 2021.

\bibitem{kwon2015roi}
S.~Kwon, J.~Kim, D.~Lee, and K.~Park.
\newblock Roi analysis for remote photoplethysmography on facial video.
\newblock In {\em 2015 37th Annual International Conference of the IEEE
  Engineering in Medicine and Biology Society (EMBC)}, pages 4938--4941. IEEE,
  2015.

\bibitem{li2013review}
C.~Li, V.~M. Lubecke, O.~Boric-Lubecke, and J.~Lin.
\newblock A review on recent advances in doppler radar sensors for noncontact
  healthcare monitoring.
\newblock {\em IEEE Transactions on microwave theory and techniques},
  61(5):2046--2060, 2013.

\bibitem{li2010comparison}
J.~Li, J.~Jin, X.~Chen, W.~Sun, and P.~Guo.
\newblock Comparison of respiratory-induced variations in photoplethysmographic
  signals.
\newblock {\em Physiological measurement}, 31(3):415, 2010.

\bibitem{magagnin2010heart}
V.~Magagnin, M.~Mauri, P.~Cipresso, L.~Mainardi, E.~N. Brown, S.~Cerutti,
  M.~Villamira, and R.~Barbieri.
\newblock Heart rate variability and respiratory sinus arrhythmia assessment of
  affective states by bivariate autoregressive spectral analysis.
\newblock In {\em 2010 Computing in Cardiology}, pages 145--148. IEEE, 2010.

\bibitem{magnon2021benefits}
V.~Magnon, F.~Dutheil, and G.~T. Vallet.
\newblock Benefits from one session of deep and slow breathing on vagal tone
  and anxiety in young and older adults.
\newblock {\em Scientific Reports}, 11(1):1--10, 2021.

\bibitem{mcduff2021camera}
D.~McDuff.
\newblock Camera measurement of physiological vital signs.
\newblock {\em ACM Computing Surveys (CSUR)}, 2021.

\bibitem{mcduff2016cogcam}
D.~J. McDuff, J.~Hernandez, S.~Gontarek, and R.~W. Picard.
\newblock Cogcam: Contact-free measurement of cognitive stress during computer
  tasks with a digital camera.
\newblock In {\em Proceedings of the 2016 CHI Conference on Human Factors in
  Computing Systems}, pages 4000--4004, 2016.

\bibitem{mikhelson2011remote}
I.~V. Mikhelson, S.~Bakhtiari, T.~W. Elmer, A.~V. Sahakian, et~al.
\newblock Remote sensing of heart rate and patterns of respiration on a
  stationary subject using 94-ghz millimeter-wave interferometry.
\newblock {\em IEEE Transactions on Biomedical Engineering}, 58(6):1671--1677,
  2011.

\bibitem{min2007study}
S.~D. Min, D.~J. Yoon, S.~W. Yoon, Y.~H. Yun, and M.~Lee.
\newblock A study on a non-contacting respiration signal monitoring system
  using doppler ultrasound.
\newblock {\em Medical \& biological engineering \& computing},
  45(11):1113--1119, 2007.

\bibitem{moraveji2012breathtray}
N.~Moraveji, A.~Adiseshan, and T.~Hagiwara.
\newblock Breathtray: augmenting respiration self-regulation without cognitive
  deficit.
\newblock In {\em CHI'12 Extended Abstracts on Human Factors in Computing
  Systems}, pages 2405--2410. 2012.

\bibitem{moraveji2011peripheral}
N.~Moraveji, B.~Olson, T.~Nguyen, M.~Saadat, Y.~Khalighi, R.~Pea, and J.~Heer.
\newblock Peripheral paced respiration: influencing user physiology during
  information work.
\newblock In {\em Proceedings of the 24th annual ACM symposium on User
  interface software and technology}, pages 423--428, 2011.

\bibitem{mueller2010jogging}
F.~Mueller, F.~Vetere, M.~R. Gibbs, D.~Edge, S.~Agamanolis, and J.~G. Sheridan.
\newblock Jogging over a distance between europe and australia.
\newblock In {\em Proceedings of the 23nd annual ACM symposium on User
  interface software and technology}, pages 189--198, 2010.

\bibitem{palatini2007heart}
P.~Palatini.
\newblock Heart rate as an independent risk factor for cardiovascular disease.
\newblock {\em Drugs}, 67(2):3--13, 2007.

\bibitem{pilz2019vector}
C.~Pilz.
\newblock On the vector space in photoplethysmography imaging.
\newblock In {\em Proceedings of the IEEE/CVF international conference on
  computer vision workshops}, pages 0--0, 2019.

\bibitem{poh2011medical}
M.-Z. Poh, D.~McDuff, and R.~Picard.
\newblock A medical mirror for non-contact health monitoring.
\newblock In {\em ACM SIGGRAPH 2011 Emerging Technologies}, pages 1--1. 2011.

\bibitem{poh2010advancements}
M.-Z. Poh, D.~J. McDuff, and R.~W. Picard.
\newblock Advancements in noncontact, multiparameter physiological measurements
  using a webcam.
\newblock {\em IEEE transactions on biomedical engineering}, 58(1):7--11, 2010.

\bibitem{porges1969respiratory}
S.~W. Porges and D.~C. Raskin.
\newblock Respiratory and heart rate components of attention.
\newblock {\em Journal of experimental psychology}, 81(3):497, 1969.

\bibitem{purves2019neurosciences}
D.~Purves, G.~J. Augustine, D.~Fitzpatrick, W.~Hall, A.-S. LaMantia, and
  L.~White.
\newblock {\em Neurosciences}.
\newblock De Boeck Sup{\'e}rieur, 2019.

\bibitem{ramirez2015anxiety}
E.~Ram{\'\i}rez, A.~R. Ortega, and G.~A.~R. Del~Paso.
\newblock Anxiety, attention, and decision making: The moderating role of heart
  rate variability.
\newblock {\em International journal of psychophysiology}, 98(3):490--496,
  2015.

\bibitem{riener2009heart}
A.~Riener, A.~Ferscha, and M.~Aly.
\newblock Heart on the road: Hrv analysis for monitoring a driver's affective
  state.
\newblock In {\em Proceedings of the 1st international conference on automotive
  user interfaces and interactive vehicular applications}, pages 99--106, 2009.

\bibitem{rowe1998heart}
D.~W. Rowe, J.~Sibert, and D.~Irwin.
\newblock Heart rate variability: Indicator of user state as an aid to
  human-computer interaction.
\newblock In {\em Proceedings of the SIGCHI conference on Human factors in
  computing systems}, pages 480--487, 1998.

\bibitem{shirbani2020effect}
F.~Shirbani, N.~Hui, I.~Tan, M.~Butlin, and A.~P. Avolio.
\newblock Effect of ambient lighting and skin tone on estimation of heart rate
  and pulse transit time from video plethysmography.
\newblock In {\em 2020 42nd Annual International Conference of the IEEE
  Engineering in Medicine \& Biology Society (EMBC)}, pages 2642--2645. IEEE,
  2020.

\bibitem{slovak2012understanding}
P.~Slov{\'a}k, J.~Janssen, and G.~Fitzpatrick.
\newblock Understanding heart rate sharing: towards unpacking physiosocial
  space.
\newblock In {\em Proceedings of the SIGCHI conference on human factors in
  computing systems}, pages 859--868, 2012.

\bibitem{starr1939studies}
I.~Starr, A.~Rawson, H.~Schroeder, and N.~Joseph.
\newblock Studies on the estimation of cardiac ouptut in man, and of
  abnormalities in cardiac function, from the heart's recoil and the blood's
  impacts; the ballistocardiogram.
\newblock {\em American Journal of Physiology-Legacy Content}, 127(1):1--28,
  1939.

\bibitem{stricker2014non}
R.~Stricker, S.~M{\"u}ller, and H.-M. Gross.
\newblock Non-contact video-based pulse rate measurement on a mobile service
  robot.
\newblock In {\em The 23rd IEEE International Symposium on Robot and Human
  Interactive Communication}, pages 1056--1062. IEEE, 2014.

\bibitem{thayer2010relationship}
J.~F. Thayer, S.~S. Yamamoto, and J.~F. Brosschot.
\newblock The relationship of autonomic imbalance, heart rate variability and
  cardiovascular disease risk factors.
\newblock {\em International journal of cardiology}, 141(2):122--131, 2010.

\bibitem{van2016robust}
M.~Van~Gastel, S.~Stuijk, and G.~de~Haan.
\newblock Robust respiration detection from remote photoplethysmography.
\newblock {\em Biomedical optics express}, 7(12):4941--4957, 2016.

\bibitem{varga2017rhythms}
S.~Varga and D.~H. Heck.
\newblock Rhythms of the body, rhythms of the brain: Respiration, neural
  oscillations, and embodied cognition.
\newblock {\em Consciousness and Cognition}, 56:77--90, 2017.

\bibitem{viola2001rapid}
P.~Viola and M.~Jones.
\newblock Rapid object detection using a boosted cascade of simple features.
\newblock In {\em Proceedings of the 2001 IEEE computer society conference on
  computer vision and pattern recognition. CVPR 2001}, volume~1, pages I--I.
  Ieee, 2001.

\bibitem{vogels2018fully}
T.~Vogels, M.~Van~Gastel, W.~Wang, and G.~De~Haan.
\newblock Fully-automatic camera-based pulse-oximetry during sleep.
\newblock In {\em Proceedings of the IEEE Conference on Computer Vision and
  Pattern Recognition Workshops}, pages 1349--1357, 2018.

\bibitem{wang2018comparative}
C.~Wang, T.~Pun, and G.~Chanel.
\newblock A comparative survey of methods for remote heart rate detection from
  frontal face videos.
\newblock {\em Frontiers in bioengineering and biotechnology}, 6:33, 2018.

\bibitem{wang2020impact}
W.~Wang and C.~Shan.
\newblock Impact of makeup on remote-ppg monitoring.
\newblock {\em Biomedical Physics \& Engineering Express}, 6(3):035004, 2020.

\bibitem{xiao2019fast}
S.~Xiao, J.~Nie, R.~Tan, X.~Duan, J.~Ma, Q.~Li, and T.~Wang.
\newblock Fast-response ionogel humidity sensor for real-time monitoring of
  breathing rate.
\newblock {\em Materials chemistry frontiers}, 3(3):484--491, 2019.

\bibitem{xiao2006frequency}
Y.~Xiao, J.~Lin, O.~Boric-Lubecke, and M.~Lubecke.
\newblock Frequency-tuning technique for remote detection of heartbeat and
  respiration using low-power double-sideband transmission in the ka-band.
\newblock {\em IEEE Transactions on Microwave Theory and Techniques},
  54(5):2023--2032, 2006.

\bibitem{zhang2010sound}
T.~Zhang, W.~Ser, G.~Y.~T. Daniel, J.~Zhang, J.~Yu, C.~Chua, and I.~Louis.
\newblock Sound based heart rate monitoring for wearable systems.
\newblock In {\em 2010 International Conference on Body Sensor Networks}, pages
  139--143. IEEE, 2010.

\bibitem{zhao2013remote}
F.~Zhao, M.~Li, Y.~Qian, and J.~Z. Tsien.
\newblock Remote measurements of heart and respiration rates for telemedicine.
\newblock {\em PloS one}, 8(10):e71384, 2013.

\end{thebibliography}

\newpage
\appendix

\section{Supplementary Material}
\begin{figure}[h]
\centering
\includegraphics[scale=0.35]{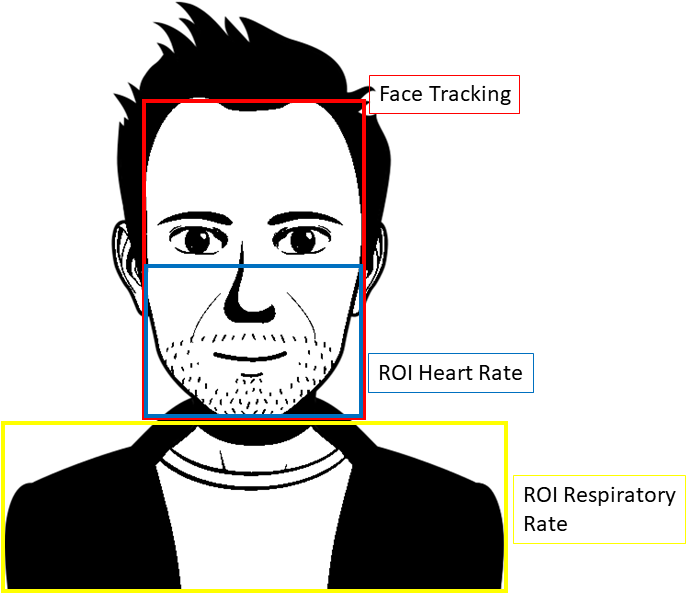}
\caption{ROI for HR and respiratory rate prediction}
\label{fig:ROI}
\end{figure}

\begin{figure}[h]
\centering
\includegraphics[scale=0.6]{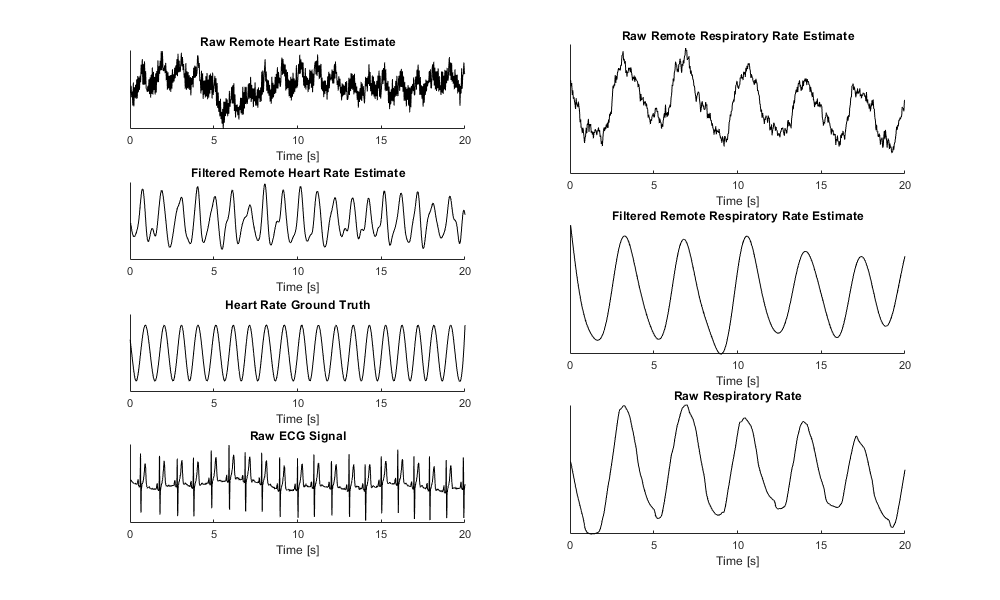}
\caption{Example of the estimation signals. On the left side the HR signals and estimates are displayed, on the right side the respiratory rate signals and estimates can be seen. The raw remote estimates corresponds hereby to the unprocessed camera data while the raw ECG and respiratory rate signals are from the ECG electrodes and respiration belt.}
\label{fig:estimate}
\end{figure}

\end{document}